Science Mapping to study academic knowledge circulation

Julián D. Cortés is a Distinguished Professor at the School of Management at Universidad Del Rosario (Colombia) and an Invited Researcher to the Fudan Development Institute (FDDI) at Fudan University, China, working in higher education since 2012.

Zaida Chinchilla-Rodríguez is an Associate Research Professor at the Spanish National Research Council (CSIC) - Institute of Public Goods and Policies (IPP) in Madrid (Spain). Member of the *SCImago Research Group* since its inception.

Katerina Bohle Carbonell is a post-doctoral researcher at the Whitaker Institute of Public Policy at the National University of Galway, Ireland. She works at the intersection of social network research, futurizing educational institution, and people management.

# Science Mapping to study academic knowledge circulation

## Abstract

The application of mathematics and statistical methods to scholarly communication: scientometrics, has facilitated the systematic analysis of the modern digital tide of literature. This chapter reviews three of such applications: co-authorship, bibliographic coupling, and co-word techniques. It also presents an exploratory case of study for the *knowledge circulation* literature. It was found a diverse geographical production, mainly in the Global North and Asian institutions with significant intermediation of universities from USA, Colombia, and Japan. The research fronts identified were related to science and medicine's history and philosophy; education, health, policy studies; and a set of interdisciplinary topics. Finally, the knowledge pillars were comprised of urban planning policy, economic geography, and historical and theoretical perspectives in the Netherlands and Central Europe; globalization and science, technology, and innovation and historical and institutional frameworks in the UK; and cultural and learning studies in the XXI century.

## Introduction

The *Oekonomische Encyklopädie* ⊠an encyclopedia on economics assembled by Johann Georg Krünitz in the XVIII century⊠ consisted of 242 volumes of 600-800 pages each (Burke, 2013). It is hard to imagine the sum of resources and effort invested at that time to publish, circulate and share the knowledge of such a collection. However, the academic publishing business has changed abruptly since the dawn of the *information age* (McGuigan & Russell, 2008). The publication's output is increasing annually at ≈9% (1946-2010), doubling its volume every ≈8 years (Bornmann & Mutz, 2015). As could be assumed, such growth and further access and adoption of academic knowledge are unevenly distributed upon the world due to countries' and regions' differences in historical, institutional and cultural pathways (Chinchilla-Rodríguez et al., 2019; Keim et al., 2014).

Bibliometrics has contributed to shedding light upon those and other explanatory factors related to scholarly communication dynamics (Chinchilla-Rodríguez et al., 2018; King, 2004; Klavans & Boyack, 2017; Larivière, Haustein, et al., 2015). Today, its methods and applications have spread to computer science, business, economics, engineering, and other research domains.

Implementations of bibliometric techniques to the specific concerns of *knowledge circulation* have been either output or citation-based studies. Using output and citation scores, Kong and Qian (2019) investigated the location of knowledge production, contributors, research fronts, and the circulation/impact of works to disentangle the "Anglo-American hegemony" in urban studies. Further citation-based studies argued that non-Anglophone ontologies are conceived as exotic stories, used to enrich the Anglophone perspective with evidence (i.e., anecdotes) (Bański & Ferenc, 2013; Wang & Zhang, 2020). The focus on output (i.e., productive) and impact (i.e., counting citation) on scholarly communication studies leave an open angle for contributing to the study of academic knowledge circulation (AKC) applying structural techniques such as science mapping (SM).

SM enables to gain insights and to depict the development of scientific disciplines or research fields on a specific topic in the current digital ocean of scholarly communication by identifying individual/institutional social capital and emerging or mature/seminal research fields, and the knowledge flows between actors (e.g., authors, institutions, papers) (Börner et al., 2012; Boyack & Klavans, 2010; Callon et al., 1983; Kessler, 1963; Shiffrin & Börner, 2004; Small, 1973). SM also visualizes the connection (i.e., edges) between actors (i.e., nodes) involved in scientific knowledge production, highlighting how academic knowledge is shared, recombined, and developed over time and space (Börner, 2010).

This chapter reviews SM techniques to analyze the structure of AKC as a research topic. It also discusses the results of applying SM techniques to the *knowledge circulation* literature as an exploratory study case. Findings would be of interest to research policy assessment entities from national to organizational levels, editorials, and research funders to understand the AKC from a quantitative angle enriched by a case of study based on a recent literature review on the topic.

The following section presents a review of the seminal studies and posterior influencing studies of three SM techniques: co-authorship, bibliographic coupling, and co-word analysis. It then shows the data sourced, methods, and software implemented for the exploratory case study. In the subsequent two sections, the results are analyzed and discussed in contrast with previous findings of SM research. Finally, the conclusion presents further lines of inquiry.

## Science mapping to study academic knowledge circulation: a review

Newman (2003) proposed four types of networks: social, information, technological, and biological. The study of AKC via SM relies upon the first two types. Van Eck and Waltman (2014) argued that all SM techniques are relation-based analyses and can be divided into three categories: citation relations, keyword co-occurrences, and co-authorships relations. In sum, direct citation, bibliographic coupling and co-citation analysis are citation-based; co-word analyses are cognitive and co-authorship analysis are social-based. Each type of analysis provides different insights. Co-authorship analysis

Co-authorship analysis (CA) is an ideal tool for mapping scientific collaboration (i.e., scientists co-creating academic knowledge). Scientific collaboration is associated with science *professionalization* and *specialties*, which defines a scientific community's rules, rights, access, differentiation, and relationships with outsiders (Watson, 2016). Such a community also sets, develops, exploits, and circulates via scholarly communication a shared body of knowledge produced reciprocally but asymmetrically by more or less influential figures (Beaver & Rosen, 1978; Chinchilla-Rodríguez et al., 2019; Edge & Mulkay, 1976).

In a CA, a node represents an author, department, institution, or country. Two or more nodes are connected through an edge if they appeared as coauthors in the same document (Figure 1). The interconnectedness between nodes has been used to estimate research impact, reveal a fluent AKC, mutual intellectual influence and affinity, and share equipment or financial resources (Hâncean & Perc, 2016; Newman, 2004; Xie et al., 2016). Any scientific collaboration that does not result in an academic publication or a given author is not included in the CA (Frassl et al., 2018).

Unpacking the issue, Sonnenwald (2007) proposes four stages of scientific collaboration: i) foundation, where scientific, political, socio-economic, resource, and previous social capital factors influence, ii) formulation, where research mission/vision and execution and intellectual property are structured, iii) sustainment, where learning and communication channels mature and emerging challenges are solved, iv) and conclusion, where collaboration success is assessed and results are disseminated.

Findings on scientific collaboration (Dong et al., 2017), stated that collaboration generated 90% of the world-leading innovations in the XXI compared to those in the 1900s. Also, it was found that international collaboration and citations between 1900 to 2015 have increased 25 and 7-fold, respectively. However, scientific collaboration is not distributed homogeneously upon the world (Chinchilla-Rodríguez et al., 2019).

Scientific collaboration among BRICS countries (i.e., Brazil, Russia, India, China, and South Africa) is weaker than their collaboration with other 65 countries (Finardi & Buratti, 2016). A relevant finding as BRICS countries agreed towards mutual thriving. However, this mutual growth requires further support. This endemic AKC can be attributed to foundation factors and geographical and idiomatic similarities (Chinchilla-Rodríguez et al., 2012).

*Authors:*

McCartney, P.[1,4], Lennon, J.[1,4], Starr, R.[1,4], Harrison, G.[1,4], & Taylor, D.[1,2,3]

*Affiliations:*

[1]University of Cambridge, Cambridge, UK
[2]California Institute of Technology (CalTech), Pasadena, EEUU
[3]Universidad del Rosario, Bogotá, Colombia
[4]University of Liverpool, Liverpool, UK

*Co-authorship network - Institutions*

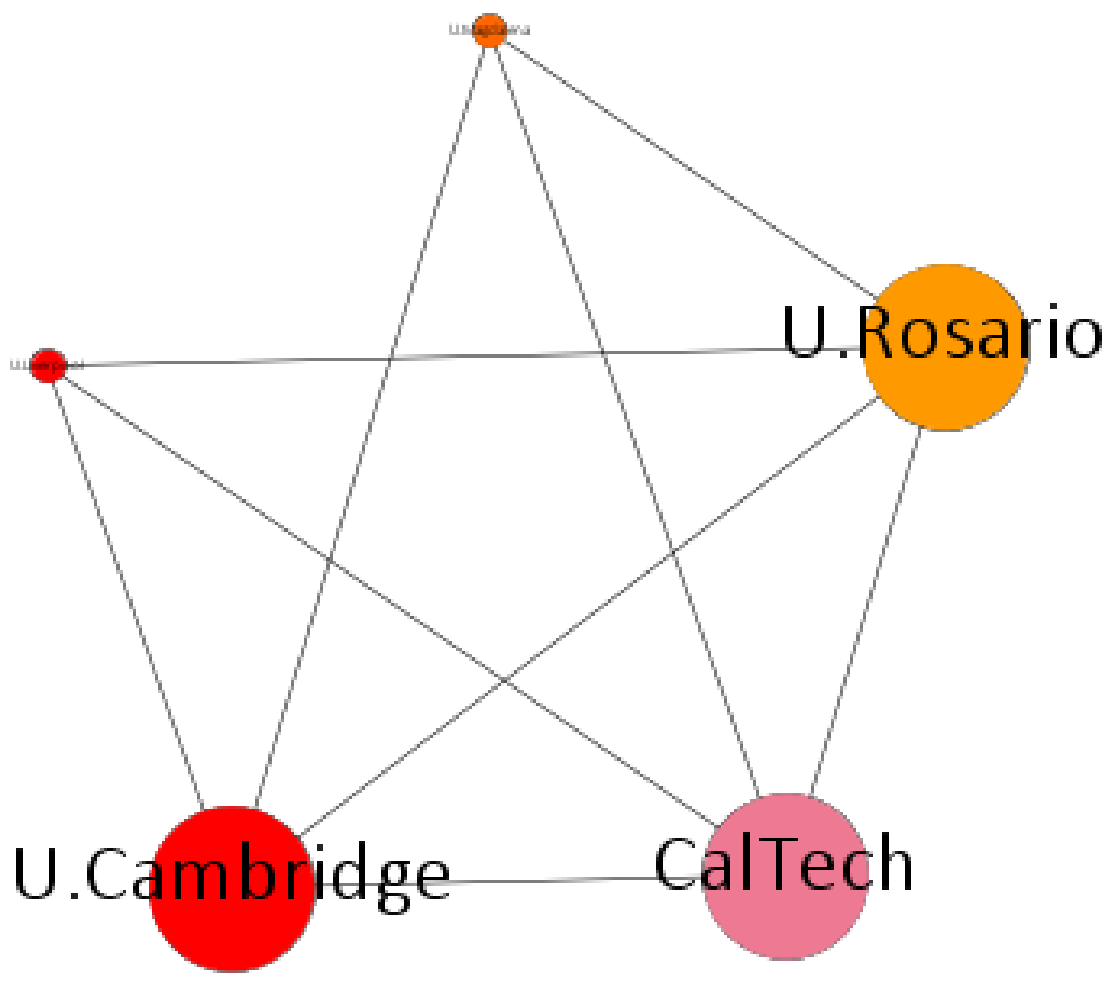

Figure 1 Example of a co-authorship network at the institutional level. Source: the authors based on Cortés-Sánchez (2020c).

*Bibliographic coupling analysis*

In the 1960s, Kessler (1963) proposed bibliographic coupling analysis (BC) to group scientific papers. The method consisted of establishing an edge between two papers (i.e., two nodes) if a common item appeared among their references (Figure 2). Kessler then outlined a few properties, such as its independence of language, the automatic classification, the groups assembled (i.e., clusters), its extension into the past and future, and the AKC changing structures according to the discipline/field emerging, ripening, or reinforcement (Weinberg, 1974). Recent refinements improved its use to analyze and visualize its clustering for investigating disciplines' hierarchy/diversity and mapping research fronts (Gómez-Núñez et al., 2014; Liu, 2017).

First, employing BC ⊠among other techniques⊠ Fanelli and Glänzel (2013) argued that the Hierarchy of the Sciences hypothesis (i.e., a set of disciplines converging theoretically and methodologically while others are diverging) is the best framework to understand disciplines' diversity ⊠so far. Second, the map of sustainability-related research highlights two major domains for the global development agenda: health and healthcare; and environment, agriculture and sustainability science (Nakamura, M; Pendlebury, D; Schnell, J; Szomszor, 2019). Global North countries dominate the landscape and Global South countries found fertile ground for collaboration with Europe.

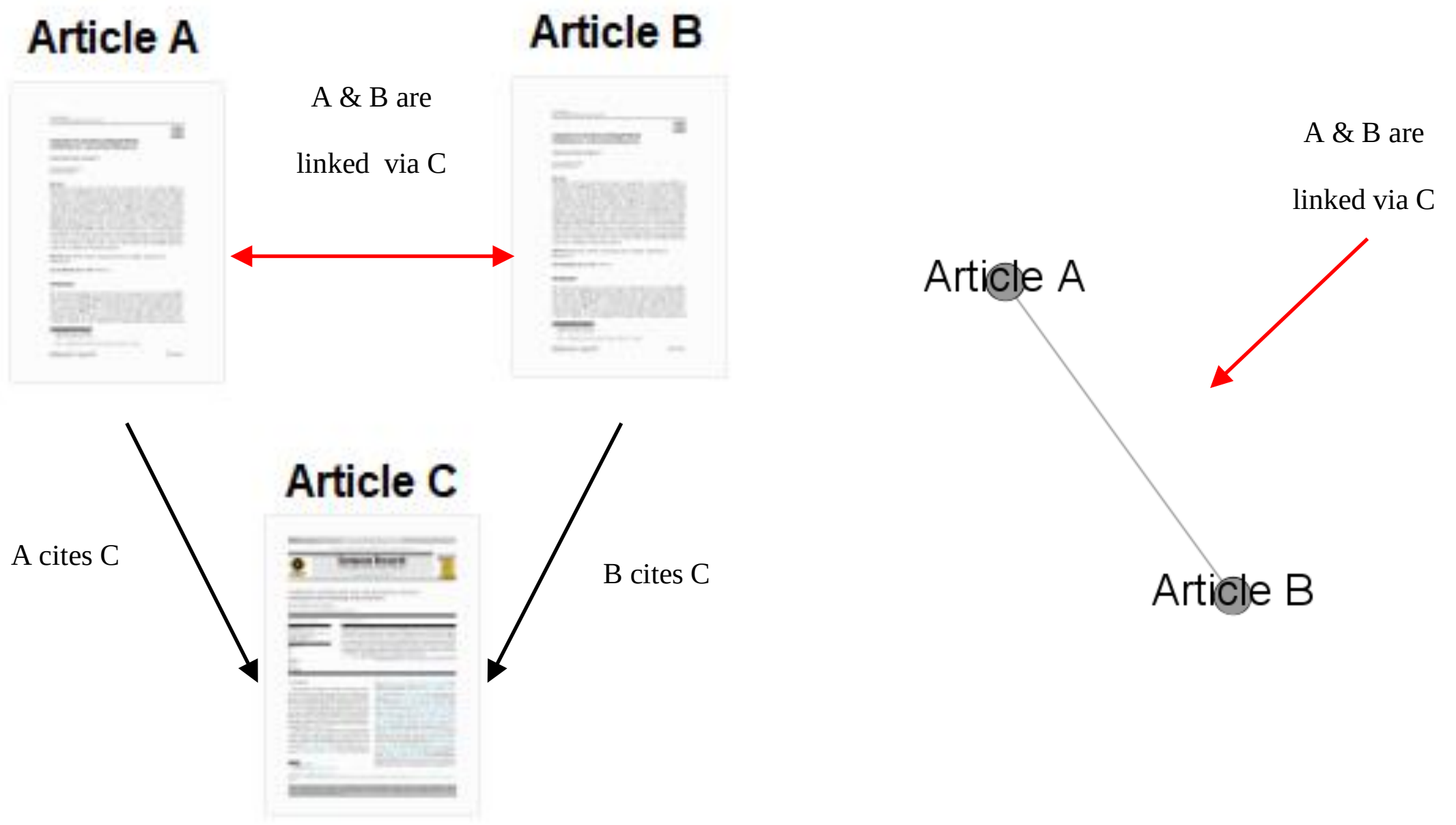

Figure 2 Example of a bibliographic coupling between two articles. Source: the authors based on Cortés-Sánchez (2020a).

*Co-word analysis*

In the early 1980s, Callon introduced the co-word analysis (CW): "Certain words and words associations are stronger than granite" (Callon et al., 1983, p. 199) Two keywords (i.e., two nodes) are linked if they appeared in the same document (Figure 3). Words were conceived as translation operators (Callon et al., 1983). The choice of words is a way for an actor to identify its relations to other actors, including the complex networks they are immersed in. Callon (1983) used a CW to develop and visualize problematic networks ⊠defining problems and putting them into a relationship with one another. The CW highlighted commonalities among problem statements and thus helps to define distinct problem space being helpful to comprehend the interaction between academic and technological research in innovation economics. Subsequent CW research restated the finding that the field of scientometrics is more similar to a *hard science* without leaving social science's realm. It also proposed a development core for the field around *scientific research evaluation* (Courtial, 1994) ⊠at that time⊠⊠

Courtial (1994) also suggested for upcoming CW and scientometrics research to embrace other disciplines, which has been effectively conducted for engineering (Coulter et al., 1998), strategic management (Ronda-Pupo & Guerras-Martin, 2012), consumer behavior (Muñoz-Leiva et al., 2012), psychophysiology (Viedma-Del-Jesus et al., 2011), and even nanoscience and nanotechnology (Muñoz-Écija et al., 2017). The CW also highlights how a journal can change its focus over time. For example, a CW conducted to commemorate the 150$^{th}$ anniversary of the journal *Nature* found that the journal started with contributions of large natural phenomena (e.g., sun, water, earth) and has been focusing on more specialized and micro grounds (e.g., protein, gene, quantum) (Monastersky & Van

Noorden, 2019) supporting the above-mentioned Hierarchy of the Sciences hypothesis observing development of converging theoretically and methodologically while focusing on smaller phenomena (Fanelli & Glänzel, 2013).

*Keywords article 1*

*sustainable development; clean production; supply chain*

*Keywords article 2*

*corporate social responsibility - CSR; sustainable development; Sustainable Development Goals - SDG*

*Keywords article 3*

*deforestation; carbón footprint; sustainable development*

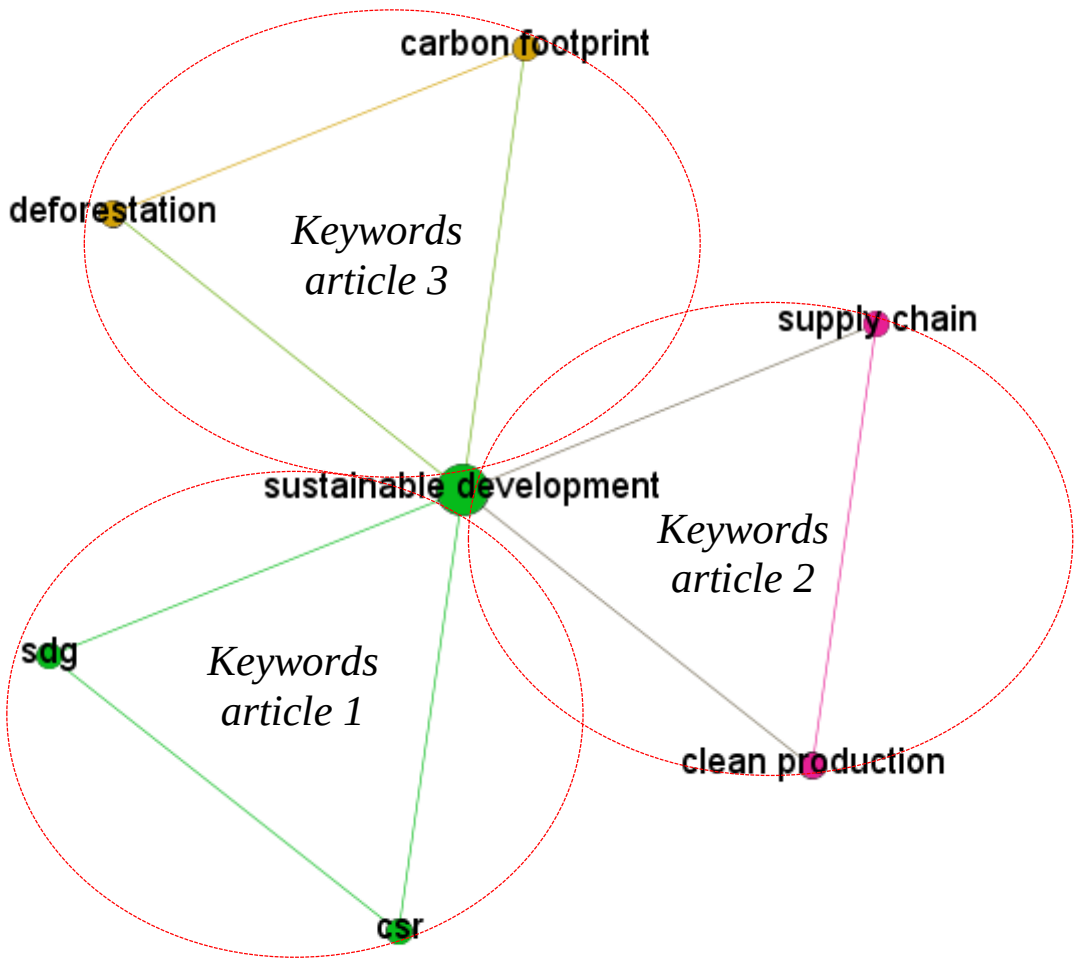

Figure 3 Example of a co-word analysis between three articles. Source: the authors based on Cortés-Sánchez (2020b).

*Citation distortions and biases*

Even though the SM techniques reviewed are based on *metaknowledge*, i.e., knowledge about knowledge (Evans & Foster, 2011), distortions and biases can be made, influencing the AKC process itself. Starting with co-authorship, the lone author was commonplace in the dawn of scientific publications; now, it is multicoauthorship (Larivière, Gingras, et al., 2015; Wuchty et al., 2007). That suggests a few inquiries on the (co)authors' role: how are contributions granted to individual authors in teams? Should the order of appearance be proportional to the author's contribution? All authors listed made a significant contribution to the research? In biomedical sciences, the junior researcher who did most of the work is attributed to the first position, while the one who supervised the work gets the last one (Casadevall et al., 2019). Others appear as authors because of their role as heads of laboratories or departments without a genuine contribution to the research (Stokes & Hartley, 1989). All the above have a further influence on researchers' careers known as the *chaperone effect:* the transition of a junior researcher into principal investigators as traced by their co-authorship position and subsequent impact (Sekara et al., 2018). Regardless of the authors' order, findings showed an overestimation of the author's contribution to the project or self-serving bias (i.e., each author self-assigned a higher contribution percentage than that assigned to them by the rest) (Herz et al., 2020).

Another bias that influences the accuracy of SM is citations. Editors and researchers have found many ways to promote their papers and raise the number of citations of their articles, the phenomenon called 'citation cartels (Fister et al., 2016). Greenberg (2011) describes several citation biases that inflate citation counts or reduce the accuracy of SM: i) *citation transmutation*, or when a hypothesis turns into a particular fact; ii) *citation diversion*, or when content is cited, but it is changed its true meaning and further implications; iii) *dead-end citation,* or when a claim is invented or leads to a non-existent source; iv) and *title invention*, or when a claim is stated in a paper's title or paragraph but is not addressed in the methodology let alone discussed in other sections. Studies published with statistically significant results are cited twice often than negative studies, controlling for research topic (Jannot et al., 2013). Consequently, these citation biases limit SM's insights since citations are one of the primary substrate of several SM techniques

However, SM contributions to study AKC are fourfold. The CA maps the individual or institutional producers and carriers of AKC, helping to identify intellectual (sometimes invisible) communities and internal hierarchies and highlighting potential collaboration and resource-sharing paths. The BC map how AKC is clustered into research fronts and their changing structures, relationships, and divergence. The CW map networks in a (multi/trans/inter)disciplinary way and root/seminal topics.

## Mapping the academic knowledge circulation literature as a case of study

### *Data*

Data was retrieved from Scopus (2020). Scopus was chosen over Web of Science (WoS) due to its broader journal coverage to conduct large-scale studies for science policy assessment at a national, institutional, and individual level (Baas et al., 2020). The query for research articles with the key-term *knowledge circulation* or *circulation of knowledge* on titles, abstracts, or keywords found 514 documents, among articles (60.7%), reviews (16.1%), conference papers (10.5%), and books (3.5%), written by 782 authors between 1991 and 2019.

We focus our analysis on the subsample composed of 312 articles, after removing two. That size reaches the benchmark for a *body of knowledge* exploration, which rounds 80 documents (Desrochers et al., 2016). duplicates using the Damerau-Levenshtein distance at 95%. The final sample was composed of 310 articles. Most of them were published by authors affiliated with institutions from the USA (9%), France (8%), Japan (7%), Germany (7%), and the UK (6%).

### *Methods and software*

Three scores at the macro, meso, and micro levels will be used to showcase SM's usefulness for analysis and interpretation. Density and average path length scores were used to describe macro features of the AKC network. Density indicates how connected the field is, and average path length represents the efficiency of AKC (Brandes, 2001; Iacobucci et al., 2018). Density measures the number of connections that exist compared to the number of potential connections that could exist. Average path lengths indicate how long, on average, an interaction chain is from one network node to another network node.

Meso scores indicate a community type of structure (i.e., modularity) and how many communities are found (i.e., clusters' number) (Blondel et al., 2008; Traag et al., 2019). Modularity measures a networks' cluster division strength. In short, the higher the modularity, the denser the connections among nodes within clusters, and the sparse connection between nodes in other clusters.

Betweenness is a micro score that unveils nodes' capacity to enhance AKC between clusters, a kind of intellectual bridge between communities (Opsahl et al., 2010). To conduct the analysis, the R packages (R Core Team, 2014) Bibliometrix (Aria & Cuccurullo, 2017) and igraph (The igraph core team, 2019), were used. Gephi (Bastian M. et al., 2009) was also essential for NA indices and networks' layout. The networks' layout was generated using the Fruchterman-Reingold algorithm (Fruchterman & Reingold, 1991). It simulates the network as a system of mass particles. Therefore, the nodes (i.e., institutions, keywords, documents) are the particles and the edges (i.e., co-authorships, keywords co-occurrences, indirect citations) are *springs* between them, pulling those connected towards each other and repelling those dispersed from each other.

Three segments compose each SM results section. First, in a top-down fashion, it will present the macro (density and average path length), meso (modularity and number of clusters), and micro (betweenness centrality) scores. It will then zoom into the largest communities (i.e., clusters) and plausible labeling. Third, it will show its intellectual/social bridges.

## Results and discussion

### *Co-authorship analysis*

Figure 4 presents the CA, composed of 203 institutions. Its density score was 0.01, which means that the network did not reach more than 2% of *real* connections out of the proportion of *potential* connections. The average path length of the networks assembled (i.e., 1.04) means that a researcher(s) affiliated with any institution would have to contact one institution on average to reach other researcher(s) affiliated with a third institution. In other words, there is, on average, only one "middle man." The CA had modularity of 0.95 and is composed of 74 clusters. That indicates a high level of collaboration within clusters, with shallow interaction between clusters. The largest cluster of the co-authorship network (pink=5.91% institutions) was geographically diverse, containing universities from North America (e.g., Stanford, Cornell, and Toronto), Europe (e.g., C*entre d'Etudes et de Recherche en Gestion d'Aix-Marseille,* Bocconi, and Hildesheim) and Asia (e.g., East China Normal). The second-largest cluster (green=3.45%) was entirely composed of USA universities (i.e., Indiana, Rice, MIT). The following cluster (blue=2.96%) was primarily European and Brazilian (i.e., French National Center for Scientific Research, Royal Holloway, Brasilia, and Federal University of Pará). In essence, the research on *knowledge circulation* has been published and circulated among Global North and Asian institutions, with Brazil's marginal involvement. Finally, at the micro level, the CA has a few but geographically diverse institutions with high betweenness scores (Table I), such as Standford (USA), Externado (Colombia), and Kyoto (Japan). These universities act as social bridges between other institutions.

The diverse scientific collaboration depicted in the largest cluster and the average path length of the CA supports Dong et al. (2017) and Wagner et al. (2015) findings related to the steady increase and multicontinental co-authorship ⊠⊠particularly with an emerging global powerhouse such as China (Tollefson, 2018). On the other hand, the *brokerage* role of Stanford University could be described by Sonnenwald's (2007) stages. Such position could be influenced by its institutional foundation, formulation, sustainment, or conclusion factors. Universities with betweenness scores (e.g., Externado [Colombia] and Kyoto [Japan]) could profit from such factors, although not at the same level. Either way, the multiple continental *brokers* identified set a position regarding institutional enablers of the AKC geographically located outside the "Anglo-American hegemony," as stated by Kong and Qian (2019). In other words, different institutions from those in the Global North have a strategic role in the AKC of the *knowledge circulation* research. Furthermore, that *brokerage* and leading position was concentrated in a few institutions since only five out of 203 institutions showed some betweenness scores, as Chinchilla-Rodríguez et al. (2019) argued.

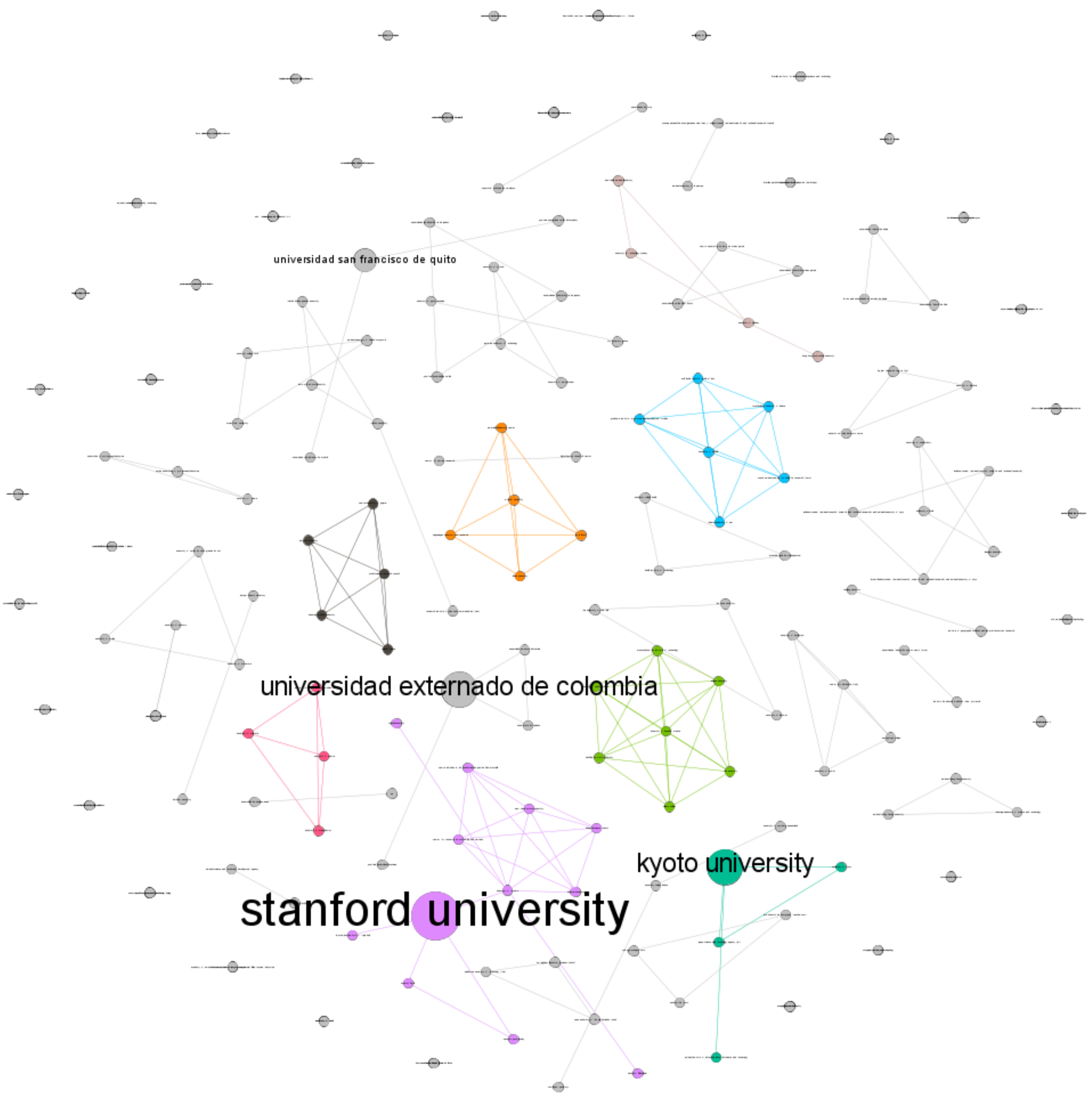

Figure 4 Co-authorship network. Source: the author based on Scopus (2020) and processed with Bibliometrix (Aria & Cuccurullo, 2017) and Gephi (Bastian et al., 2009). Layout algorithm: Fruchterman-Reingold (1991). Note: node and label size proportional to its betweenness score.

*Bibliographic coupling analysis*

Figure 5 presents the BC. It was composed of 302 documents. It had a density of 6.7% and its average path length was 2.46. That means that a researcher/department working on a specific topic (e.g., institutional framework) would have to gain expertise in more than two different topics or body of knowledge on average to reframe its focus in any other different one either from the same (e.g., historical geography) or different cluster (e.g., urban policy). The BC had a modularity score of 0.38 and they were detected 53 clusters. While the largest number of clusters could indicate a high fragmentation of the field, the low level of modularity suggests that these clusters are penetrable.

The BC highlighted three research fronts. The largest cluster (pink=27.48% documents) refers to science and medicine's history and philosophy. The second and third research fronts were the green one (18.21%), encompassing education, health, policy studies, and blue one (14.24%), covering a range of interdisciplinary topics, from renaissance studies to environmental planning and ecology.

The intellectual bridge (Table I) was that article of Hall (2009): *"Ecologies of business education and the geographies of knowledge."* That article was nested in the fuchsia cluster (4.3% of documents). It aimed to "*develop geographical perspectives on knowledge and learning to situate business education within broader landscapes of corporate knowledge circulation, production and learning*" (Hall, 2009, p. 599). This work's higher betweenness score indicates that it was grounded in the body of knowledge of multiple and cognitive-distant research fronts, ranging from cultural change and future studies to education, health, and policy studies.

The BC clustered the research on *knowledge circulation* orbiting social sciences and humanities (e.g., science and medicine's history and philosophy). However, the BC of *knowledge circulation* research was more similar in terms of modularity and average path length to those of psychiatry/psychology (modularity: 0.48; average path length: 2.9) than those of social sciences (modularity: 0.62; average path length: 3.9) and arts and humanities (modularity: 0.79; average path length: 5.7) calculated by Fanelli and Glänzel's (2013) BC to investigate the Hierarchy of the Sciences hypothesis. Despite such discordance, results reaffirm that the *knowledge circulation* research methods and theories will likely diverge. The highly disciplinary diverse article of Hall (2009) (i.e., the intellectual bridge of the BC) is a testimony on that. There are also potential matches with the global development agenda and its main cluster (i.e., health and healthcare) considering the second most populated cluster here assembled (i.e., education and health policy) (Nakamura, M; Pendlebury, D; Schnell, J; Szomszor, 2019). Even management and business research has been permeated by the sustainable development agenda (Cortés-Sánchez, 2019, 2020b; Cortés-Sánchez et al., 2020).

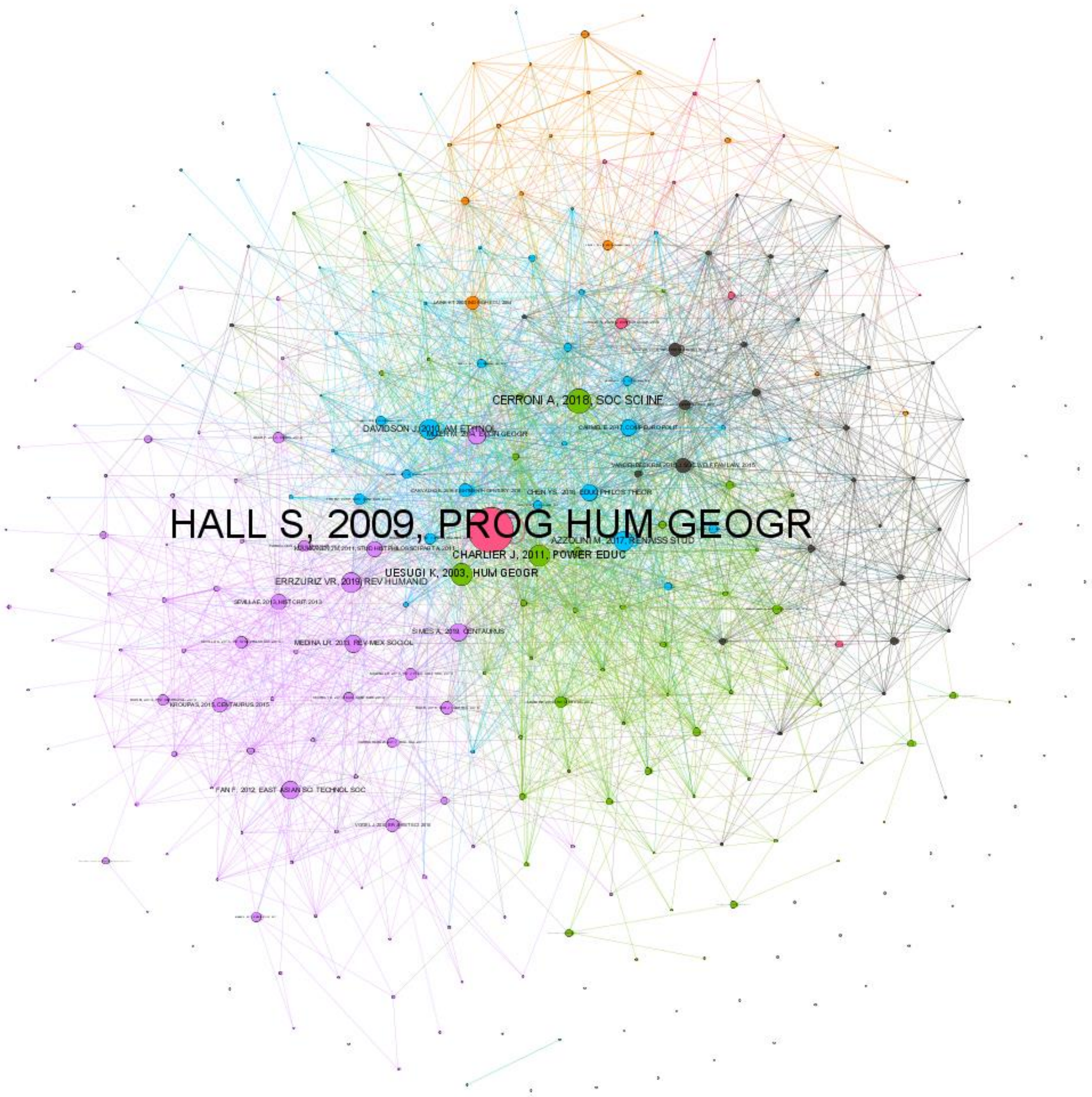

Figure 5 Bibliographic coupling network. Source: the author based on Scopus (2020) and processed with Bibliometrix (Aria & Cuccurullo, 2017) and Gephi (Bastian et al., 2009). Layout algorithm: Fruchterman-Reingold (1991). Note: node and label size proportional to its betweenness score.

### *Co-word analysis*

The CW was composed of 724 key-terms with a density score of 0.014 (Figure 6). The average path length was 3.32 with a modularity score of 0.66 and 16 clusters. The higher modularity score indicates stronger borders between clusters and fewer bridges/links/connections between them. That indicates that while theoretical frameworks and empirical insights are more easily shared among research fronts, they are embedded in a group-specific language/jargon. The CW largest cluster (pink=6.63% key-terms) dealt with urban planning policy, economic geography, and historical and theoretical perspectives in the Netherlands and Central Europe. The second-largest cluster (green=6.49%) was composed of topics related to globalization and science, technology, and innovation and historical and institutional frameworks in the UK. The third cluster (blue=6.35%) dealt with cultural and learning studies in the XXI century. Moving to the intellectual bridges (Table I), the CW highlighted studies about/from France, human(s), and perception.

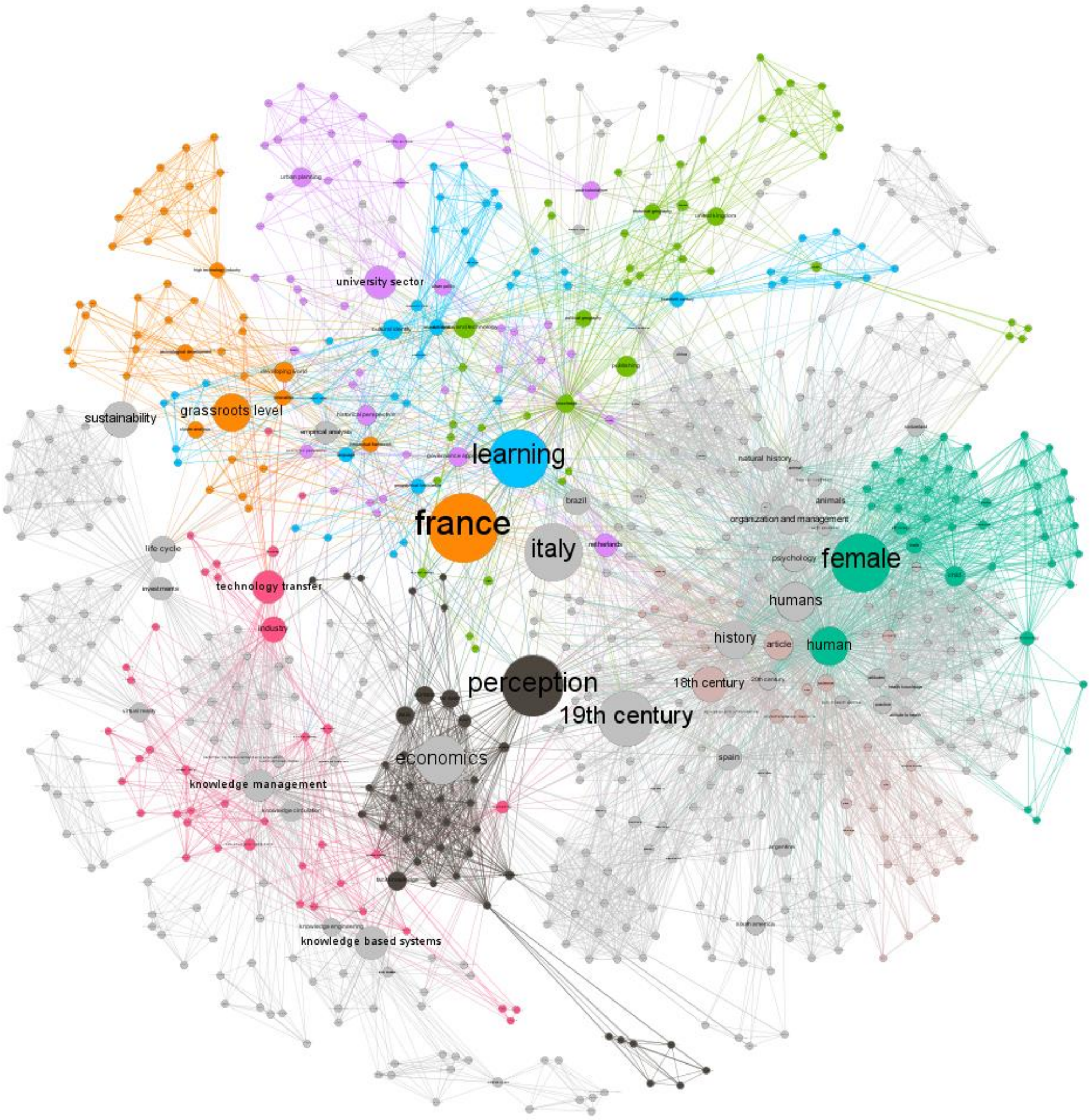

Figure 6 Co-word network. Source: the author based on Scopus (2020) and processed with Bibliometrix (Aria & Cuccurullo, 2017) and Gephi (Bastian et al., 2009). Layout algorithm: Fruchterman-Reingold (1991). Note: node and label size proportional to its betweenness score.

Table I Top-five institutions, key terms and documents with the highest betweenness scores

| | CA | | BC | | CW | |
|---|---|---|---|---|---|---|
| # | Institution | Betweenness | Document | Betweenness | Key term | Betweenness |
| 1 | Stanford University (USA) | 0.000074 | Hall S, 2009, *Prog Hum Geogr* | 0.055042 | France | 0.017987 |
| 2 | Universidad Externado (Colombia) | 0.000049 | Cerroni A, 2018, *Soc Sci Inf* | 0.028389 | Humans | 0.01797 |
| 3 | Kyoto University (Japan) | 0.000049 | Uesugi K, 2003, *Hum Geogr* | 0.025217 | Perception | 0.015506 |
| 4 | Universidad San Francisco de Quito (Ecuador) | 0.000025 | Charlier J, 2011, *Power Educ* | 0.025087 | Female | 0.014787 |
| 5 | | | Errzuriz Vr, 2019, *Rev Humanid* | 0.022774 | Learning | 0.014757 |

Source: the author based on Scopus (2020) and processed with Bibliometrix (Aria & Cuccurullo, 2017) and Gephi (Bastian et al., 2009).

## Conclusion

We reviewed the potential of combining data mining and visualization SM techniques to understand better. Such SM techniques enabled us to map scientific collaboration, research fronts, and research-problems space. The CA showed a high and diverse level of collaboration, mainly in the Global North. The BC highlighted three research fronts related to science and medicine; education, health, policy studies; and interdisciplinary topics. For its part, the CW also identified the three largest clusters on urban planning, economic geography, and theoretical perspectives in the Netherlands and Central Europe.

Research policy entities from national to organizational levels, editorials, and research funders could use the SM here presented to elaborate more or and focused research agendas and funding programs, directed to distribute resources to either mature or emergent fields among the AKC literature. Also, to promote collaboration between the protagonist and peripherical institutions.

Every methodological approach has limitations, so does SM. For example, further research could use broader and inclusive bibliographic databases such as Google Scholar or Dimensions and refine the search strategy. Besides, as a limited range of tools and analysis types has been used in this chapter compared to the suite of freely available scientometrics mapping tools, we planned to extend our research considering all the above-mentioned limitations via more refined valuations (e.g., at the individual author-researcher level). Finally, identifying key funding actors and stakeholders interested in the AKC domain would guide us to address specific research questions to help in gaining a high–level understanding of the evolution and development of AKC.

## Acknowledgments

[Pending]